\documentstyle[aps,pre,twocolumn,epsfig,floats]{revtex}

\newcommand{\equ}[1]{(\protect\ref{#1})}
\begin{document}
\draft 
\wideabs{ 
\title{Epidemic dynamics and endemic states in complex networks}
 
\author{Romualdo Pastor-Satorras$^1$,  
and Alessandro Vespignani$^2$}

\address{$^1$Dept. de F\'{\i}sica i Enginyeria Nuclear,
  Universitat Polit\`{e}cnica de Catalunya,\\
  Campus Nord, M\`{o}dul B4,  08034 Barcelona, Spain   \\
  $^2$The Abdus Salam International Centre 
  for Theoretical Physics (ICTP),
  P.O. Box 586, 34100 Trieste, Italy\\
}
\date{\today}
\maketitle

\begin{abstract}
  We study by analytical methods and large scale simulations a
  dynamical model for the spreading of epidemics in complex networks.
  In networks with exponentially bounded connectivity we recover the
  usual epidemic behavior with a threshold defining a critical point
  below which the infection prevalence is null.  On the contrary, on a
  wide range of scale-free networks we observe the absence of an
  epidemic threshold and its associated critical behavior. This
  implies that scale-free networks are prone to the spreading and the
  persistence of infections whatever spreading rate the epidemic
  agents might possess. These results can help understanding computer
  virus epidemics and other spreading phenomena on communication and
  social networks.
\end{abstract}
 
}

\section{Introduction}

Many social, biological, and communication systems can be properly
described by complex networks whose nodes represent individuals or
organizations and links mimic the interactions among them
\cite{net1,amaral}.  Recently, many authors have recognized the
importance of local clustering in complex networks. This implies that
some {\em special} nodes of the network posses a larger probability to
develop connections pointing to other nodes.  Particularly interesting
examples of this kind of behavior are found in metabolic networks
\cite{jeong00}, food webs \cite{ricard}, and, most importantly, in the
Internet and the world-wide-web, where the networking properties have
been extensively studied because of their technological and economical
relevance \cite{amaral,int,www,www2}.

Complex networks can be classified in two main groups, depending on
their connectivity properties. The first and most studied one is
represented by the {\em exponential} networks, in which the nodes'
connectivity distribution (the probability $P(k)$ that a node is
connected to other $k$ nodes) is exponentially bounded
\cite{rg,er,ws}.  The typical example of an exponential network is the
random graph model of Erd\"{o}s and R\'{e}nyi \cite{er}. A network
belonging to this class that has recently attracted a great deal of
attention is the Watts and Strogatz model (WS) \cite{ws,amaral2,alain}, which
has become the prototypical example of a {\em small-world} network
\cite{watts99}.  A second and very different class of graph is
identified by the {\em scale-free} (SF) networks which exhibit a
power-law connectivity distribution \cite{barab99},
\begin{equation}
  P(k)\sim k^{-2-\gamma},
  \label{pgamma}
\end{equation}
where the parameter $\gamma$ must be larger than zero to ensure a
finite average connectivity $\left< k \right>$.  This kind of
distribution implies that each node has a statistically significant
probability of having a very large number of connections compared to
the average connectivity of the network $\left< k \right>$. In
particular, we will focus here on the Barab\'{a}si and Albert model
(BA) \cite{barab99}, which results in a connectivity distribution
$P(k)\sim k^{-3}$.

In view of the wide occurrence of complex networks in nature, it
becomes a very interesting issue to inspect the effect of their
complex features on the dynamics of epidemic and disease
spreading\cite{mn99}, and more in general on the nonequilibrium phase
transitions that usually characterize these type of phenomena
\cite{reviews}. It is easy to foresee that the characterization and
understanding of epidemic dynamics on these networks can find
immediate applications to a wide range of problems, ranging from
computer virus infections \cite{vir}, epidemiology \cite{biomodels},
and the spreading of polluting agents \cite{pol}.

In this paper, we shall study the susceptible-infected-susceptible
(SIS) model \cite{biomodels} on complex networks.  We study
analytically the prevalence and persistence of infected 
individuals in exponential and
SF networks by using a single-site approximation that takes into
account the inhomogeneity due to the connectivity distribution. We
find that exponential networks show, as expected, an epidemic
threshold (critical point) separating an infected from a healthy
phase. The density of infected nodes decreases to zero at the
threshold with the linear behavior typical of a mean-field (MF)
critical point \cite{reviews}. The SF networks, on the other hand,
show a very different and surprising behavior. For $0<\gamma\leq 1$
the model does not show an epidemic threshold and the infection can
always pervade the whole system. In the region $1<\gamma\leq 2$, the
model shows an epidemic threshold that is approached, however, with a
vanishing slope; i.e.  in the absence of critical fluctuations. Only
for $\gamma>2$ we recover again the usual critical behavior at the
threshold.  In these systems, because of the nonlocal connectivity,
single site approximation predictions are expected to correctly depict
the model's behavior. In order to test our predictions, we perform
large scale numerical simulations on both exponential and SF networks.
Numerical results are in perfect agreement with the analytical
predictions and confirm the overall picture for the SIS model on
complex networks given by the theoretical analysis.  The striking
absence of an epidemic threshold on SF networks, a characteristic
element in mathematical epidemiology, radically changes many of the
conclusions drawn in classic epidemic modeling. The present results
could be relevant also in the field of absorbing-phase transitions and
catalytic reactions in which the spatial interaction of the reactants
can be modeled by a complex network \cite{reviews}.

The paper is organized as follows. In Sec. II we introduce the SIS
model in a general context. Sec. III is devoted to the analysis of
exponentially bounded networks, exemplified by the WS model. In Sec.
IV we analyze the scale-free BA model, with connectivity $P(k)\sim
k^{-3}$. Sec. V extends the analytical approach applied to the BA
model to generalized SF networks, with connectivity distribution
$P(k)\sim k^{-2-\gamma}$, $\gamma > 0$.  Finally, in Sec.  VI we draw
our conclusions and perspectives.

\section{The SIS model}

To address the effect of the topology of complex networks in epidemic
spreading we shall study the standard susceptible-infected-susceptible
(SIS) epidemiological model \cite{biomodels}.  Each node of the
network represents an individual and each link is a connection along
which the infection can spread to other individuals.  The SIS model
relies on a coarse grained description of the individuals in the
population.  Within this description, individuals can only exist in
two discrete states, namely, susceptible, or ``healthy'', and
infected.  These states completely neglect the details of the
infection mechanism within each individual.  The disease transmission
is also described in an effective way.  At each time step, each
susceptible node is infected with probability $\nu$ if it is connected
to one or more infected nodes. At the same time, infected nodes are
cured and become again susceptible with probability $\delta$, defining
an effective spreading rate $\lambda=\nu/\delta$.  (Without lack of
generality, we set $\delta=1$.) Individuals run stochastically through
the cycle susceptible $\to$ infected $\to$ susceptible, hence the name
of the model.  The updating can be performed with both parallel or
sequential dynamics\cite{reviews}.  The SIS model does not take into
account the possibility of individuals removal due to death or
acquired immunization\cite{biomodels}. It is mainly used as a
paradigmatic model for the study of infectious disease that leads to
an endemic state with a stationary and constant value for the
prevalence of infected individuals, i.e., the degree to which the
infection is widespread in the population.

The topology of the network that specifies the interactions among
individuals is of primary importance in determining many of the
model's features. In standard topologies the most significant result
is the general prediction of a nonzero epidemic threshold
$\lambda_c$~\cite{biomodels}.  If the value of $\lambda$ is above the
threshold, $\lambda\geq \lambda_c$, the infection spreads and becomes
persistent in time. Below it, $\lambda < \lambda_c$, the infection
dies out exponentially fast. In both sides of the phase diagram it is
possible to study the behavior in time of interesting dynamical
magnitudes of epidemics, such as the time survival probability and the
relaxation to the healthy state or the stationary endemic state.  In
the latter case, if we start from a localized seed we can study the
epidemic outbreak preceding the endemic stabilization.  From this
general picture, it is natural to consider the epidemic threshold as
completely equivalent to a critical point in a nonequilibrium phase
transition \cite{reviews}.  In this case, the critical point separates
an active phase with a stationary density of infected nodes (an
endemic state) from an absorbing phase with only healthy nodes and
null activity.  In particular, it is easy to recognize that the SIS
model is a generalization of the contact process (CP) model, that has
been extensively studied in the context of absorbing-state phase
transitions\cite{reviews}.

In order to obtain an analytical understanding of the SIS model
behavior on complex networks, we can apply a single site dynamical 
MF approach, that we
expect to recover exactly the model's behavior due to the nonlocal
connectivity of these graphs. Let us consider separately the case of
the exponentially bounded and SF networks.

\section{Exponential networks: the Watts-Strogatz  model}

The class of exponential networks refers to random graph models which
produce a connectivity distribution $P(k)$ peaked at an average value
$\left<k\right>$ and decaying exponentially fast for $k \gg
\left<k\right>$ and $k\ll\left<k\right>$.  Typical examples of such a
network are the random graph model\cite{er} and the small-world model
of Watts and Strogatz (WS) \cite{ws}. The latter has recently been the
object of several studies as a good candidate for the modeling of many
realistic situations in the context of social and natural networks. In
particular, the WS model shows the ``small-world'' property common in
random graphs \cite{watts99}; i.e., the diameter of the graph--- the
shortest chain of links connecting any two vertices---increases very
slowly, in general logarithmically with the network size \cite{alain}.
On the other hand, the WS model has also a local structure (clustering
property) that is not found in random graphs with finite
connectivity\cite{ws,alain}.  The WS graph is defined as follows
\cite{ws,alain}: The starting point is a ring with $N$ nodes, in which
each node is symmetrically connected with its $2K$ nearest neighbors.
Then, for every node each link connected to a clockwise neighbor is
rewired to a randomly chosen node with probability $p$, and preserved
with probability $1-p$. This procedure generates a random graph with a
connectivity distributed exponentially for large $k$ \cite{ws,alain},
and an average connectivity $\left<k \right> = 2 K$. The graphs has
small-world properties and a non-trivial ``clustering coefficient'';
i.e., neighboring nodes share many common neighbors \cite{ws,alain}.
The richness of this model has stimulated an intense activity aimed at
understanding the network's properties upon changing $p$ and the
network size $N$~\cite{ws,amaral2,alain,watts99,nw99,bra99}. At the
same time, the behavior of physical models on WS graphs has been
investigated, including epidemiological percolation
models\cite{mn99,nw99,call00} and models with epidemic
cycles\cite{abra00}.

Here we focus on the WS model with $p=1$; it is worth noticing that
even in this extreme case the network retains some memory of the
generating procedure. The network, in fact, is not locally equivalent
to a random graph in that each node has at least $K$ neighbors.  Since
the fluctuations in the connectivity are very small in the WS graph,
due to its exponential distribution, we can approach the analytical
study of the SIS model by considering a single MF reaction equation
for the density of infected nodes $\rho(t)$:
\begin{equation}
\partial_t \rho(t) = -\rho(t) +\lambda \left< k \right> 
\rho(t) \left[ 1-\rho(t) \right] +h.o.~.
\label{eq:ws}
\end{equation}
The MF character of this equation stems from the fact that we have
neglected the density correlations among the different nodes,
independently of their respective connectivities.  In Eq.~\equ{eq:ws}
we have ignored all higher order corrections in $\rho(t)$, since we
are interested in the onset of the infection close to the phase
transition; i.e., at $\rho(t) \ll 1$.  The first term on the r.h.s. in
Eq.~\equ{eq:ws} considers infected nodes become healthy with unit
rate. The second term represents the average density of newly infected
nodes generated by each active node. This is proportional to the
infection spreading rate $\lambda$, the number of links emanating from
each node, and the probability that a given link points to a healthy
node, $\left[ 1-\rho(t) \right]$.  In these models, connectivity has
only exponentially small fluctuations ($\left< k^2 \right> \sim
\left<k\right>$) and as a first approximation we have considered that
each node has the same number of links, $k\simeq \left< k \right>$.
This is equivalent to an homogeneity assumption for the system's
connectivity.  After imposing the stationarity condition $\partial_t
\rho(t) =0$, we obtain the equation
\begin{equation}
  \rho \left[ -1 + \lambda \left< k \right> (1-\rho) \right]=0
\end{equation}
for the steady state density $\rho$ of infected nodes.  This equation
defines an epidemic threshold $\lambda_c=\left< k \right>^{-1}$, and
yields:
\begin{mathletters}
\begin{eqnarray}
  \rho &=& 0 \;\qquad\qquad \mbox{\rm if $\lambda< \lambda_c$}, \\
  \rho &\sim& \lambda-\lambda_c \qquad \mbox{\rm if
  $\lambda>\lambda_c$} \label{eq:meanfield}.
\end{eqnarray}
\end{mathletters}
In analogy with critical phenomena, we can consider $\rho$ as the
order parameter of a phase transition and $\lambda$ as the tuning
parameter, recovering a MF critical behavior \cite{nota3}. It is
possible to refine the above calculations by introducing connectivity
fluctuations (as it will be done later for SF networks, see Sec. IV).
However, the results are qualitatively and quantitatively the same as
far as we are only concerned with the model's behavior close to the
threshold.

In order to compare with the analytical prediction we have performed
large scale simulations of the SIS model in the WS network with $p=1$.
Simulations were implemented on graphs with number of nodes ranging
from $N=10^3$ to $N=3 \times 10^6$, analyzing the stationary
properties of the density of infected nodes $\rho$; i.e. the infection
prevalence. Initially we infect half of the nodes in the network, and
iterate the rules of the SIS model with parallel updating. In the
active phase, after an initial transient regime, the systems stabilize
in a steady state with a constant average density of infected nodes.
The prevalence is computed averaging over at least $100$ different
starting configurations, performed on at least $10$ different
realization of the random networks. In our simulations we consider the
WS network with parameter $K=3$, which corresponds to an average
connectivity $\left< k \right> = 6$.

\begin{figure}[t]
  \centerline{\epsfig{file=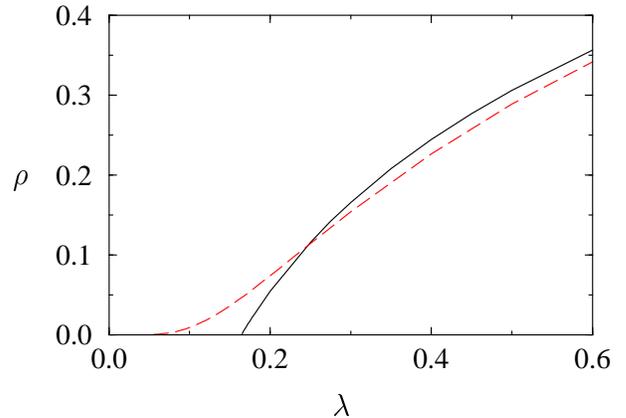, width=8cm}}
  \caption{Density of infected nodes $\rho$ as a function of
    $\lambda$ in the WS network (full line) and the BA network (dashed
    line).}
  \label{fig:fig1}
\end{figure}

\begin{figure}[t]
  \centerline{\epsfig{file=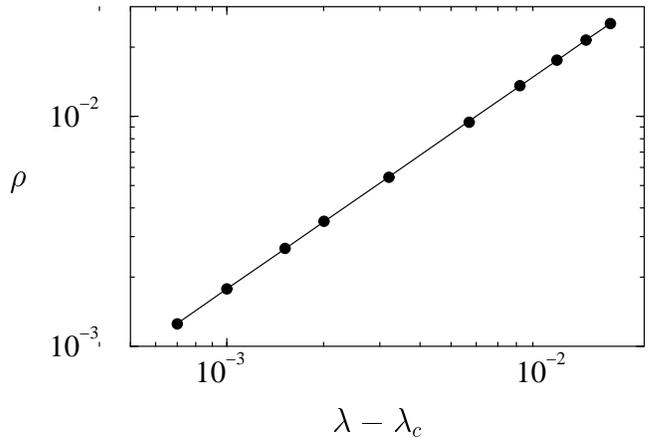, width=8.5cm}}
  \caption{Log-log plot of density of infected node $\rho$ as a
    function of $\lambda-\lambda_c$ in WS network, with
    $\lambda_c=0.1643\pm0.01$. The full line is a fit to the form
    $\rho \sim (\lambda-\lambda_c)^\beta$, with an exponent
    $\beta=0.97\pm0.04$.}
  \label{fig:fig2}
\end{figure}

As shown in Figs.~\ref{fig:fig1} and~\ref{fig:fig2}, the SIS model on
a WS graph exhibits an epidemic threshold $\lambda_c=0.1643\pm0.01$
that is approached with linear behavior by $\rho$. The value of the
threshold is in good agreement with the MF predictions $\lambda_c=1/2
K = 0.1666$, as well as the density of infected nodes behavior.  In
Fig.~\ref{fig:fig2} we plot $\rho$ as a function of
$\lambda-\lambda_c$ in log-log scale. A linear fit to the form $\rho
\sim (\lambda-\lambda_c)^\beta$ provides an exponent
$\beta=0.97\pm0.04$, in good agreement with the analytical finding 
of the Eq.~\equ{eq:meanfield}.

\begin{figure}[t]
  \centerline{\epsfig{file=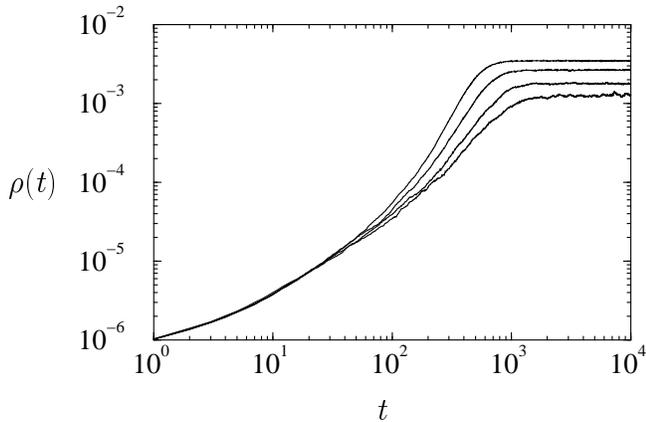, width=8.5cm}}
  \caption{Density of infected nodes $\rho(t)$ as a function of
    time in supercritical spreading experiments in the WS network.
    Network size $N=1.5 \times 10^6$. Spreading rates range from
    $\lambda -\lambda_c= 0.002$ to $0.0007$ (top to bottom).}
  \label{fig:fig3}
\end{figure}

\begin{figure}[t]
  \centerline{\epsfig{file=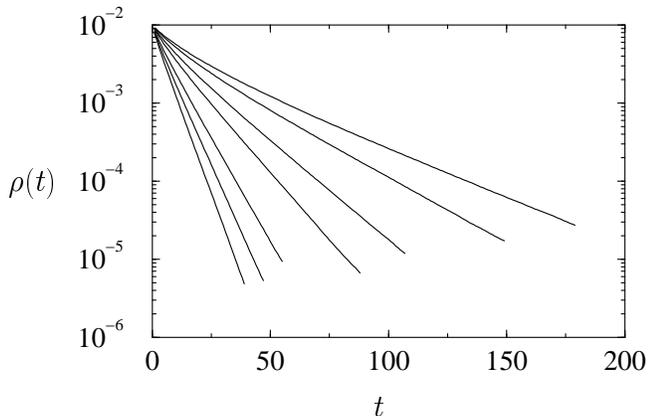, width=8.5cm}}
  \caption{Density of infected nodes $\rho(t)$ as a function of
    time in subcritical spreading experiments in the WS network.
    Network size $N=3 \times 10^6$. Spreading rates range from
    $\lambda_c-\lambda = 0.005$ to $0.03$ (right to left).}
  \label{fig:fig4}
\end{figure}

\begin{figure}[t]
  \centerline{\epsfig{file=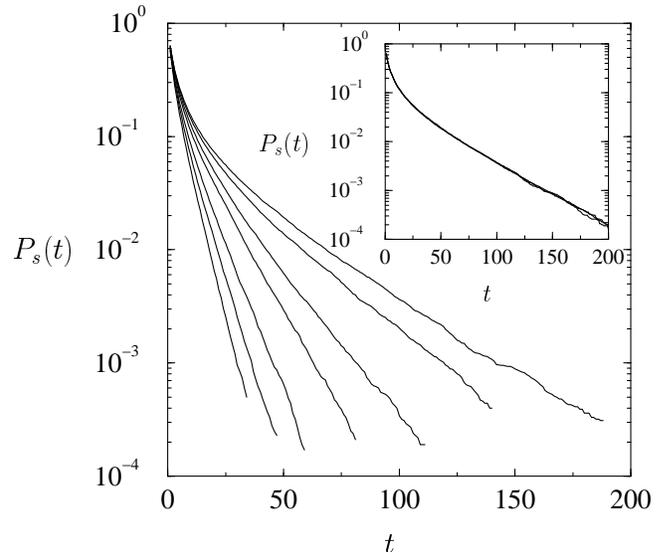, width=8.5cm}}
  \caption{Surviving probability $P_s(t)$ as a function of
    time in subcritical spreading experiments in the WS network.
    Network size $N=3 \times 10^6$. Spreading rates range from
    $\lambda_c-\lambda = 0.005$ to $0.03$ (right to left). Inset:
    Surviving probability for a fixed spreading rate
    $\lambda_c-\lambda = 0.005$ and network sizes $N=3 \times 10^5$,
    $10^6$, and $3 \times 10^6$.}
  \label{fig:fig5}
\end{figure}

\begin{figure}[t]
  \centerline{\epsfig{file=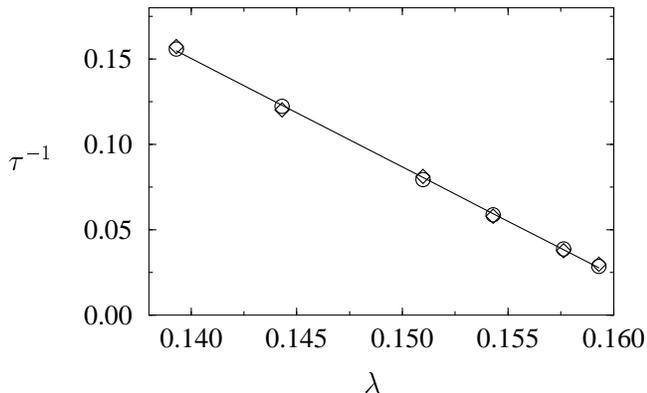, width=8.5cm}}
  \caption{Inverse relaxation time for the SIS model in the WS graph
    as a function of the spreading rate $\lambda$, estimated from the
    slope of the exponential decay of the infected nodes density
    $\rho(t)$ ($\circ$) and the survival probability $P_s(t)$
    ($\Diamond$).}
  \label{fig:fig6}
\end{figure}

To complete our study of the SIS model in the WS network, we have also
analyzed the epidemic spreading properties, computed by considering
the time evolution of infections starting from a very small
concentration of infected nodes. In Fig.~\ref{fig:fig3} we plot the
evolution of the infected nodes density as a function of time for
epidemics in the supercritical regime ($\lambda> \lambda_c$) which
start from a single infected node. Each curve represents the average
over several spreading events with the same $\lambda$.  We clearly
notice a spreading growth faster than any power-law, in agreement with
Eq.~\equ{eq:ws} which predicts an exponential saturation to the
endemic steady state.  In the subcritical regime ($\lambda <
\lambda_c$), by introducing a small perturbation to the stationary
state $\rho=0$, and keeping only first order terms in
Eq.~(\ref{eq:ws}), we obtain that the infection decays following the
exponential relaxation $\partial_t \rho(t) = -\left< k
\right>(\lambda_c-\lambda)\rho(t)$.  This equation introduces a
characteristic relaxation time
\begin{equation}
  \tau^{-1} =\left< k \right> (\lambda_c-\lambda),
  \label{eq:tau}
\end{equation}
that diverges at the epidemic threshold.  Below the threshold, the
epidemic outbreak dies within a finite time; i.e. it does not reach a
stationary endemic state. In Fig.~\ref{fig:fig4} we plot the average
of $\rho(t)$ for epidemics starting with an initial concentration
$\rho_0=0.01$ of infected nodes; the Figure shows a clear exponential
approach to the healthy (absorbing) state as predicted by
Eq.~(\ref{eq:tau}). In the subcritical regime, we can compute also the
surviving probability, $P_s(t)$, defined as the probability that an
epidemic outbreak survives up to the time $t$~\cite{reviews}. In
Fig.~\ref{fig:fig5} we plot the survival probability computed from
simulations starting with a single infected node in a WS graph of size
$N=3 \times 10^6$.  The survival probability decay is obviously
governed by the same exponential behavior and characteristic time of
the density of infected nodes as confirmed by numerical simulations.
Indeed, below the epidemic threshold, the relaxation to the absorbing
state does not depend on the network size $N$ (see inset in
Fig.~\ref{fig:fig5}), and the average lifetime corresponding to each
spreading rate $\lambda$ can be measured by the slope of the
exponential tail of $P_s(t)$ and $\rho(t)$.  By plotting $\tau^{-1}$
as a function of $\lambda_c-\lambda$ (see Fig.~\ref{fig:fig6}), we
recover the analytic predictions ; i.e. the linear behavior and the
unique characteristic time for both the density and survival
probability decay.  The slope of the graph, measured by means of a
least squares fitting, provides a value of $6.3$, whereas the
intercept yields $1.0$, in good agreement with the theoretical
predictions of Eq.~\equ{eq:tau}, $\left< k \right> = 6$ and $\left< k
\right> \lambda_c =1$, respectively.

In summary, numerical and analytical results confirms that for WS
graphs, the standard epidemiological picture (often called the
deterministic approximation) is qualitatively and quantitatively
correct. This result, that is well-known for random graphs, holds also
in the WS model despite the different local structure.

\section{Scale-free networks: the Barab\'{a}si-Albert model}

The Barab\'{a}si-Albert (BA) graph was introduced as a model of
growing network (such as the Internet or the world-wide-web) in which
the successively added nodes establish links with higher probability
pointing to already highly connected nodes \cite{barab99}.  This is a
rather intuitive phenomenon on the Internet and other social networks,
in which new individuals tend to develop more easily connections with
individuals which are already well-known and widely connected.  The BA
graph is constructed using the following algorithm \cite{barab99}: We
start from a small number $m_0$ of disconnected nodes; every time step
a new vertex is added, with $m$ links that are connected to an old
node $i$ with probability
\begin{equation}
  \Pi(k_i) = \frac{k_i}{\sum_j k_j},
\end{equation}
where $k_i$ is the connectivity of the $i$-th node.  After iterating
this scheme a sufficient number of times, we obtain a network composed
by $N$ nodes with connectivity distribution $P(k) \sim k^{-3}$ and
average connectivity $\left<k \right> = 2 m$ (in this work we will
consider the parameters $m_0=5$ and $m=3$). Despite the well-defined
average connectivity, the scale invariant properties of the network
turns out to play a major role on the properties of models such as
percolation\cite{call00,hav00}, used to mimic the resilience 
to attacks of a network. For this class of graphs, in fact, the absence 
of a characteristic scale for the connectivity makes highly connected 
nodes statistically significant, and induces strong fluctuations in the
connectivity distribution which cannot be neglected.  In order to take
into account these fluctuations, we have to relax the homogeneity
assumption used for exponential networks, and consider the relative
density $\rho_k(t)$ of infected nodes with given connectivity $k$;
i.e., the probability that a node with $k$ links is infected. The
dynamical MF reaction rate equations can thus be written as
\begin{equation}
  \partial_t \rho_k(t) = -\rho_k(t) +\lambda k \left[
  1-\rho_k(t) \right] \Theta(\rho(t)), 
\end{equation}
where also in this case we have considered a unitary recovery rate and 
neglected higher order terms ($\rho(t)\ll 1$).
The creation term considers the probability that a node with $k$ links
is healthy $[1-\rho_k(t)]$ and gets the infection via a
connected node.  The probability of this last event is proportional
to the infection rate $\lambda$, the number of connections $k$, and
the probability $\Theta(\rho(t))$ that any given link points to an
infected node. We make the assumption that $\Theta$ is a function of the
total density of infected nodes $\rho(t)$ \cite{nota4}. 
In the steady (endemic) state, $\rho$ is just a  function of
$\lambda$. Thus, the probability $\Theta$ becomes also an implicit
function of the spreading rate, and by imposing stationarity 
[$\partial_t \rho_k(t) =0$], we obtain 
\begin{equation}
  \rho_k=\frac{k \lambda\Theta(\lambda)}{1+k \lambda\Theta(\lambda)}.
  \label{nhom}
\end{equation}
This set of equations show that the higher the node connectivity, the
higher the probability to be in an infected state. This inhomogeneity
must be taken into account in the computation of $\Theta(\lambda)$.
Indeed, the probability that a link points to a node with $s$ links is
proportional to $sP(s)$.  In other words, a randomly chosen link is
more likely to be connected to an infected node with high
connectivity, yielding the relation
\begin{equation}
  \Theta(\lambda)=\sum_k \frac{kP(k)\rho_k}{\sum_s sP(s)}.
  \label{first}
\end{equation}
Since $\rho_k$ is on its turn a function of $\Theta(\lambda)$, we
obtain a self-consistency equation that allows to find
$\Theta(\lambda)$ and an explicit form for Eq.~(\ref{nhom}).  Finally,
we can evaluate the order parameter (persistence) $\rho$ using the
relation
\begin{equation}
  \rho=\sum_kP(k)\rho_k,
  \label{second}
\end{equation}
In order to perform an explicit calculation for the BA model, we use a
continuous $k$ approximation that allows the practical substitution of
series with integrals\cite{barab99}.  The full connectivity
distribution is given by $P(k)= 2m^2/k^{-3}$, where $m$ is the minimum
number of connection at each node.  By noticing that the average
connectivity is $\left< k \right> = \int_m^\infty k P(k) d k =2m$,
Eq.~(\ref{first}) gives
\begin{equation}
  \Theta(\lambda)=m\lambda\Theta(\lambda)\int_m^\infty
  \frac{1}{k^3}\frac{k^2}{1+k \lambda\Theta(\lambda)},
\end{equation}
which yields the solution
\begin{equation}
  \Theta(\lambda)=\frac{e^{-1/m\lambda}}{\lambda
  m}(1-e^{-1/m\lambda})^{-1}. 
\end{equation}
In order to find the behavior of the density of infected nodes we have
to solve Eq.~(\ref{second}), that reads as 
\begin{equation}
  \rho=2m^2\lambda\Theta(\lambda)\int_m^\infty
  \frac{1}{k^3}\frac{k}{1+k \lambda\Theta(\lambda)}.
\end{equation}
By substituting the obtained expression for $\Theta(\lambda)$ and
solving the integral we find at the lowest order in $\lambda$
\begin{equation}
\rho \sim e^{-1/m\lambda}
\end{equation}

This result shows the surprising absence of any epidemic threshold or
critical point in the model; i.e., $\lambda_c=0$. This can be
intuitively understood by noticing that for usual lattices and MF
models, the higher the node's connectivity, the smaller the epidemic
threshold. In the BA network the unbounded fluctuations in the number
of links emanating from each node ($\left< k^2 \right> =
\infty$) plays the role of an infinite connectivity, annulling thus
the threshold. This implies that infections can pervade a BA network,
whatever the infection rate they have.

\begin{figure}[t]
  \centerline{\epsfig{file=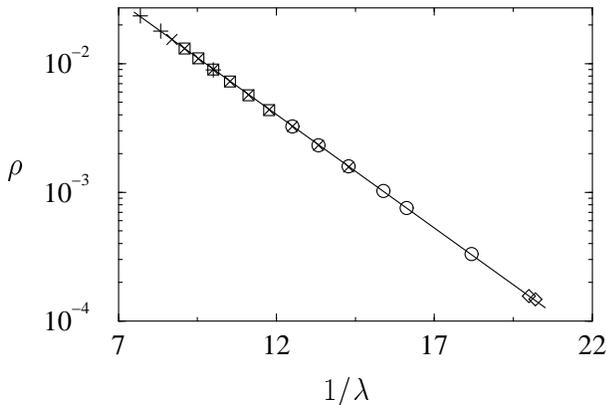, width=8cm}}
  \caption{Persistence $\rho$ as a function of $1/\lambda$ for
    BA networks of different sizes: $N=10^5$ ($+$), $N=5 \times 10^5$
    ($\Box$), $N=10^6$ ($\times$), $N=5 \times 10^6$ ($\circ$), and
    $N=8.5 \times 10^6$ ($\Diamond$). The linear behavior on the
    semi-logarithmic scale proves the stretched exponential behavior
    predicted for the persistence. The full line is a fit to the form
    $\rho \sim\exp(-C/\lambda)$.}
  \label{fig:fig7}
\end{figure}

\begin{figure}[t]
  \centerline{\epsfig{file=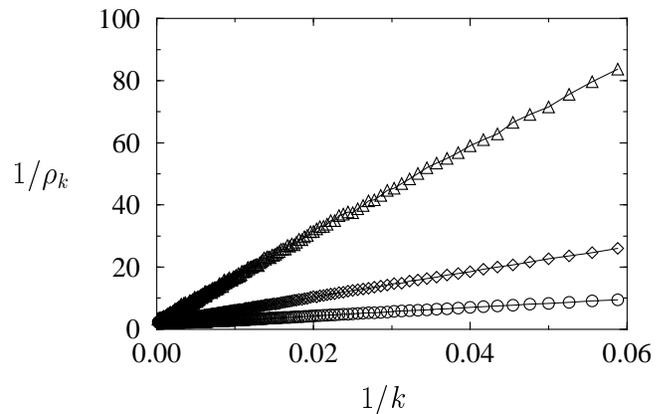, width=8.5cm}}
  \caption{The density $\rho_k$, defined as the fraction 
    of nodes with connectivity $k$ which are infected, in a BA network
    of size $N=5 \times 10^5$ and spreading rates $\lambda=0.1$,
    $0.08$ and $0.065$ (bottom to top).  The plot recovers the form
    predicted in Eq.(\ref{nhom}).}
  \label{fig:fig8}
\end{figure}

The numerical simulations performed on the BA network confirm the
picture extracted from the analytic treatment. We consider 
the SIS model on BA networks of
size ranging from $N=10^3$ to $N=8.5 \times 10^6$. In
Fig.~\ref{fig:fig1} we have plotted the epidemic persistence $\rho$ as
a function of $\lambda$ in a linear scale. The function $\rho$ 
approaches smoothly the value $\lambda=0$  with 
vanishing slope.  Fig.~\ref{fig:fig7}, in fact, shows that the
infection prevalence in the steady state decays with 
$\lambda$ as $\rho\sim \exp(-C/\lambda)$, where $C$ is a constant. The
numerical value obtained $C^{-1}=2.5$ is also in good agreement
with the theoretical prediction $C^{-1}=m=3$.  In order to rule out
the presence of finite size effect hiding an abrupt transition (the
so-called smoothing out of critical points\cite{reviews}), we have
inspected the behavior of the stationary persistence for network sizes
varying over three orders of magnitude.  The total absence of scaling
of $\rho$ and the perfect agreement for any size with the analytically
predicted exponential behavior allows us to definitely confirm the
absence of any finite epidemic threshold.  In Fig.~\ref{fig:fig8}, we
also provide an illustration of the behavior of the probability
$\rho_k$ that a node with given connectivity $k$ is infected. Also in
this case we found a behavior with $k$ in complete agreement with the
analytical prediction of Eq.~(\ref{nhom}).

\begin{figure}[t]
  \centerline{\epsfig{file=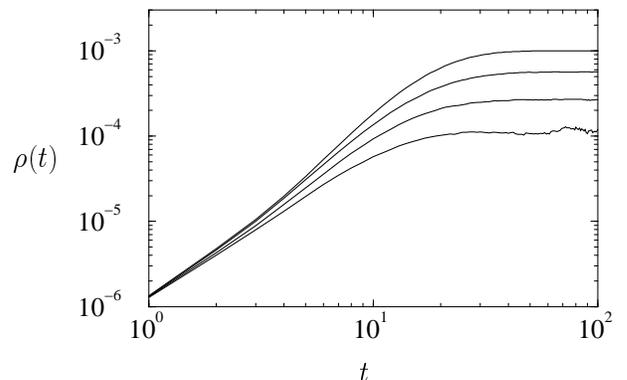, width=8cm}}
  \caption{Density of infected nodes $\rho(t)$ as a function of
    time in supercritical spreading experiments in the BA network.
    Network size $N=10^6$.  Spreading rates range from $\lambda=0.05$
    to $0.065$ (bottom to top).}
  \label{fig:fig9}
\end{figure}

\begin{figure}[t]
  \centerline{\epsfig{file=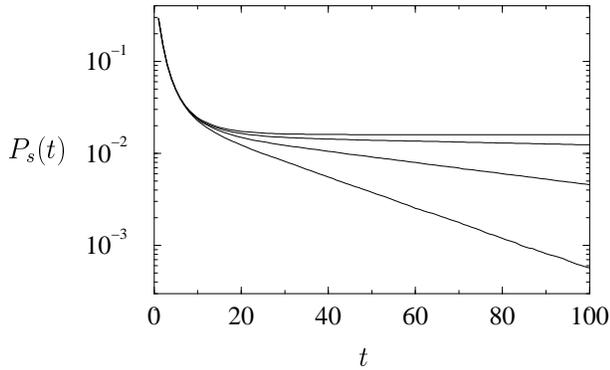, width=8cm}}
  \caption{Surviving probability $P_s(t)$ as a function of
    time in subcritical spreading experiments in the BA network.
    Spreading rate $\lambda=0.065$. Network sizes ranging from $N=6.25
    \times 10^3$ to $N=5 \times 10^5$ (bottom to top).}
  \label{fig:fig10}
\end{figure}

Our numerical study of the spreading dynamical properties on the BA
network is reported in Figs.~\ref{fig:fig9} and~\ref{fig:fig10}.  In
Fig.~\ref{fig:fig9} we plot the growth of the epidemics starting from
a single infected node. We observe that the spreading growth in time
has an algebraic form, as opposed to the exponential growth typical of
mean-field continuous phase transitions close to the critical 
point \cite{reviews}, and the behavior of the SIS model in the WS graph (see
Fig.~\ref{fig:fig3}).  The surviving probability $P_s(t)$ for a fixed
value of $\lambda$ and networks of different size $N$ is reported in
Fig~\ref{fig:fig10}. In this case, we recover an exponential behavior
in time, that has its origin in the finite size of the network.  In
fact, for any finite system, the epidemic will eventually die out
because there is a finite probability that all individuals cure the
infection at the same time.  This probability is decreasing with the
system size and the lifetime is infinite only in the thermodynamic
limit $N\to\infty$.  However, the lifetime becomes virtually infinite
(the metastable state has a lifetime too long for our observation
period) for enough large sizes that depend upon the spreading rate
$\lambda$.  This is a well-known feature of the survival probability
in finite size absorbing-state systems poised above the critical
point.  In our case, this picture is confirmed by numerical
simulations that shows that the average lifetime of the survival
probability is increasing with the network size for all the values of
$\lambda$. Given the intrinsic dynamical nature of scale-free
networks, this result could possibly have several practical
implications in the study of epidemic spreading in real growing
networks.

The numerical analysis supports and confirms the analytical results
pointing out the existence of a different epidemiological framework
for SF networks. The absence of an epidemic threshold, a central
element in the theory of epidemics, opens a different scenario and
rationalization for epidemic events in these networks. The dangerous
absence of the epidemic threshold, that leaves SF networks completely
disarmed with respect to epidemic outbreaks, is fortunately balanced
from a corresponding exponentially low prevalence at small spreading
rates. In addition, the absence of a critical threshold, and the
associated diverging response function, makes the increase of the
endemic prevalence with the spreading rate very slow. This new
perspective seems to be particularly relevant in the rationalization
of epidemic data from computer virus infections\cite{psv}.

\section{Generalized scale-free networks}

Recently there has been a burst of activity in the modeling of SF
complex network. The recipe of Barab\'{a}si and Albert\cite{barab99} has
been followed by several variations and generalizations
\cite{barab00,dogo00,krap00,bosa00} and the revamping of previous
mathematical works\cite{simon55}.  All these studies propose methods
to generate SF networks with variable exponent $\gamma$.  The
analytical treatment presented in the previous section for the SIS
model can be easily generalized to SF networks with connectivity
distribution with $\gamma>0$. Consider a generalized SF network with a
normalized connectivity distribution given by
\begin{equation}
  P(k) = (1+\gamma) m^{1+\gamma}  k^{-2-\gamma},
\end{equation}
where we are approximating the connectivity $k$ as a continuous
variable and assuming $m$ the minimum connectivity of any node.
The average connectivity is thus
\begin{equation}
  \left< k \right> = \int_m^\infty k P(k) dk = 
  \frac{1+\gamma}{\gamma}  m.    
\end{equation}
For any connectivity distribution, the relative density of infected
nodes $\rho_k$ is given by Eq.~\equ{nhom}. Applying then
Eq.~\equ{first} to compute self-consistently the probability $\Theta$, we
obtain 
\begin{equation}
  \Theta(\lambda) = F[1, \gamma, 1+\gamma, -(m \lambda
  \Theta(\lambda))^{-1}],
  \label{thetagen}
\end{equation}
where $F$ is the Gauss hypergeometric function \cite{abramovitz}. On
the other hand, the expression for the density $\rho$,
Eq.~\equ{second}, yields
\begin{equation}
  \rho=  F[1, 1+\gamma, 2+\gamma, -(m \lambda
  \Theta(\lambda))^{-1}].
  \label{rhogen}
\end{equation}

In order to solve Eqs.~\equ{thetagen} and~\equ{rhogen} in the limit
$\rho \to 0$ (which obviously corresponds also to $\Theta \to 0$, we
must perform a Taylor expansion of the hypergeometric function. The
expansion for Eq.~\equ{thetagen} has the form \cite{abramovitz}
\begin{eqnarray}
  \lefteqn{F[1, \gamma, 1+\gamma, -(m \lambda
    \Theta(\lambda))^{-1}]} \nonumber \\
  &\simeq&  \frac{\gamma \pi}{\sin(\gamma \pi)}  (m\lambda
  \Theta)^\gamma 
  + \gamma \sum_{n=1}^\infty (-1)^n
    \frac{(m\lambda\Theta)^n}{n-\gamma}, 
    \label{expansion}
\end{eqnarray}
where $\Gamma(x)$ is the standard gamma function.  An analogous
expression holds for \equ{rhogen}. The expansion \equ{expansion} is
valid for any $\gamma \neq 1, 2, 3, \ldots$. Integer values of
$\gamma$ must be analyzed in a case by case basis. (The particular
value $\gamma=1$ was considered in the previous Section.) For all
values of $\gamma$, the leading behavior of Eq.~\equ{rhogen} is the
same:
\begin{equation}
  \rho \simeq \frac{1+\gamma}{\gamma} m \lambda \Theta
  \label{leadingrho}
\end{equation}
The leading behavior in the r.h.s. of Eq.~\equ{expansion}, on the
other hand, depends of the particular value of $\gamma$:

(i) $0<\gamma<1$: In this case, one has
\begin{equation}
  \Theta(\lambda) \simeq \frac{\gamma \pi}{\sin(\gamma \pi)} 
  (m\lambda  \Theta)^\gamma,
\end{equation}
from which we obtain
\begin{equation}
  \Theta(\lambda) \simeq \left[\frac{\gamma \pi}{\sin(\gamma \pi)}
  \right]^{1/(1-\gamma)} 
  \left( m\lambda \right)^{\gamma/(1-\gamma)}. 
\end{equation}
Combining this with Eq.~\equ{leadingrho}, we obtain:
\begin{equation}
  \rho \sim \lambda^{1/(1-\gamma)}.
\end{equation}
Here we have again the total absence of any epidemic threshold and the
associated critical behavior, as we have already shown for the case
$\gamma=1$. In this case, however, the relation between $\rho$ and
$\lambda$ is given by a power law with exponent $\beta=
1/(1-\gamma)$; i.e., $\beta>1$. 

(ii) $1<\gamma<2$: In this case, to obtain a nontrivial information
for $\Theta$, we must keep the first two most relevant terms in
Eq.~\equ{expansion}: 
\begin{equation}
  \Theta(\lambda) \simeq \frac{\gamma \pi}{\sin(\gamma \pi)} (m\lambda 
  \Theta)^\gamma + \frac{\gamma}{\gamma-1} m\lambda
  \Theta.
\end{equation}
From here we get:
\begin{equation}
  \Theta(\lambda) \simeq \left[ \frac{-\sin(\gamma
    \pi)}{\pi(\gamma-1)} 
    \frac{m}{(m\lambda)^\gamma} 
    \left( \lambda - \frac{\gamma -1}{m \gamma}\right)
  \right]^{1/(\gamma-1)}.
\end{equation}
The  expression for $\rho$ is finally:
\begin{equation}
  \rho \simeq \left( \lambda - \frac{\gamma -1}{m
  \gamma}\right)^{1/(\gamma-1)} \sim \left(\lambda -
   \lambda_c
  \right)^{1/(\gamma-1)}.
\end{equation}
That is, we obtain a power-law behavior with exponent
$\beta=1/(\gamma-1)>1$, but now we observe the presence of a nonzero
threshold
\begin{equation}
  \lambda_c = \frac{\gamma -1}{m \gamma}.
  \label{threshold}
\end{equation}
In this case, a critical threshold reappears in the model. However, 
the epidemic threshold is approached smoothly without any sign of 
the singular behavior associated to critical point. 
 
(iii) $\gamma>2$: The relevant terms in the $\Theta$ expansion are now 
\begin{equation}
  \Theta(\lambda) \simeq 
  \frac{\gamma}{\gamma-1} m\lambda \Theta - 
  \frac{\gamma}{\gamma-2} (m\lambda \Theta)^2,
\end{equation}
and the relevant expression for $\Theta$ is
\begin{equation}
  \Theta(\lambda) \simeq 
  \frac{\gamma-2}{\gamma-1} \frac{1}{\lambda^2 m} \left(\lambda -
  \frac{\gamma -1}{m \gamma}\right)
\end{equation}
which yields the behavior for $\rho$
\begin{equation}
  \rho \sim \lambda - \lambda_c
\end{equation}
with the same threshold $\lambda_c$ as in Eq.~\equ{threshold} and an
exponent $\beta=1$.  In other words, we recover the usual critical
framework in networks with connectivity distribution that decays
faster than $k$ to the fourth power. Obviously, an exponentially
bounded network is included in this last case, recovering the results
obtained with the homogeneous approximation of Sec. III.

In summary, for all SF networks with $0<\gamma\leq 1$, we recover the
absence of an epidemic threshold and critical behavior; i.e.  $\rho=0$
only if $\lambda=0$, and $\rho$ has a vanishing slope when $\lambda\to
0$. In the interval $1 < \gamma < 2$, an epidemic threshold reappears
($\rho\to 0$ if $\lambda\to \lambda_c$), but it is also approached
with vanishing slope; i.e. no singular behavior.  Eventually, for
$\gamma>2$ the usual MF critical behavior is recovered and the SF
network is indistinguishable from an exponential network.

\section{Conclusions}

The emerging picture for disease spreading in complex networks
emphasizes the role of topology in epidemic modeling. In particular,
the absence of epidemic threshold and critical behavior in a wide
range of SF networks provides an unexpected result that changes
radically many standard conclusions on epidemic spreading.  Our
results indicate that infections can proliferate on these networks
whatever spreading rates they may have. These very bad news are,
however, balanced by the exponentially small prevalence for a wide
range of spreading rates ($\lambda\ll1$).  This point appears to be
particularly relevant in the case of technological networks such as
the Internet and the world-wide-web that show a SF connectivity with
exponents $\gamma\simeq 2.5$ \cite{int,www}. For instance, the present
picture fits perfectly with the observation from real data of computer
virus spreading, and could solve the longstanding problem of the
generalized low prevalence of computer viruses without assuming any
global tuning of the spreading rates\cite{vir,psv}. The peculiar
properties of SF networks also open the path to many other questions
concerning the effect of immunity and other modifications of epidemic
models. As well, the critical properties of many nonequilibrium
systems could be affected by the topology of SF networks. Given the
wide context in which SF networks appears, the results obtained here
could have intriguing implications in many biology and social systems.
 
\section*{}
\acknowledgments

This work has been partially supported by the European Network 
Contract No. ERBFMRXCT980183.  RP-S also acknowledges support from the
grant CICYT PB97-0693. We thank S.~Franz, M.-C. Miguel, R. V. Sol{\'e},
M.~Vergassola, S.~Zapperi, and R. Zecchina for comments and
discussions.


\begin{thebibliography}{10}
\bibitem{nota3}
        On Euclidean lattices the order parameter behavior is 
        $\rho\sim(\lambda-\lambda_c)^\beta$, with $\beta\leq 1$. 
        The linear behavior is recovered  above
        the upper critical dimension (see Ref.\cite{reviews}).
\bibitem{nota4}
        One could be tempted to impose  $\Theta(\rho)=\rho$;
        however, the highly inhomogeneous density $\rho_k$
        makes this approximation too strong.
\end{thebibliography}
\end{document}